\documentclass[twocolumns]{aa}
\usepackage{graphicx}
\usepackage{natbib}
\bibpunct{(}{)}{;}{a}{}{,}
\usepackage{txfonts}
\begin{document}
   \title{On the geometrical evolution of the ionized gas in HII galaxies}

   \author{F. Cuisinier
          \inst{1}
          \and
          P. Westera
          \inst{2}
          \and
          E. Telles
          \inst{3}
          \and
          R. Buser
          \inst{2}
          }

   \offprints{F. Cuisinier}

   \institute{GEMAC - Departamento de Astronomia / Observat\'orio do Valongo,
             Universidade Federal do Rio de Janeiro,
             Ladeira do Pedro Antonio 43, 20080-090 Rio de Janeiro RJ,
             Brazil
             \email{francois@ov.ufrj.br}
         \and
             Department of Physics and Astronomy,
             University of Basel, Venusstr. 7,
             CH-4102 Binningen, Switzerland
             \email {westera@astro.unibas.ch, buser@astro.unibas.ch}
         \and
             Observat\' orio Nacional/MCT, Rua Jos\'e Cristino, 77,
             Rio de Janeiro RJ, 20921-400, Brazil 
             \email{etelles@on.br}
             }

   \date{Received; Accepted}

   \abstract
            {}
            {In this paper, we investigate the behaviour of the number
             of Lyman continuum ionizing photons as compared to the actual
             number of hydrogen recombinations in HII galaxies.}
             {We evaluate
             the number of ionizing photons from the population synthesis of
             spectra observed in the visible, extrapolating the spectra to the
             extreme ultraviolet (EUV), beyond the Lyman limit. We check for possible systematic deviations of the predicted
             ionizing spectra in the EUV by comparing the ratio of the
             predicted number of ionizing photons to the number of
             recombinations, as measured in H$\beta$,
             ${\rm \Delta \log Q(H^0)}$, with the metallicity.
             We find that, as far as the number of ionizing photons is
             concerned, no systematic tendency can be detected.
%
%

              The ${\rm H\beta}$ equivalent width can be understood as a
              nebular age indicator, decreasing with  age, although the
              observed ${\rm H\beta}$ equivalent width can also be affected
              by the contribution to the continuum by the accumulation of
              previous, non-ionizing stellar populations.}
             {We attribute
              the increase of ${\rm \Delta \log Q(H^0)}$ with the age of the burst
              to the fact that more and more ionizing photons escape the
              nebulae when the nebulae get older, because of their
              increasing, expansion-induced subfragmentation.}
              {}

   \maketitle
%

\section{Introduction}
%
%
%
%

In the giant HII regions present in HII galaxies, photoionization converts
ionizing photons of the Lyman continuum in the extreme ultra-violet (EUV)
(e.g., with $\lambda \leq 912 \AA$ or E $\geq 13.6$~eV) into emission lines,
ranging from the ultra-violet to the far infrared; the most prominent ones
are recombination lines and collisional lines of atoms,
generally in their first and second ionization stages.
For photons in the EUV, the radiative
world is very different from the one for photons with
higher wavelengths. In the latter,  small amounts of neutral hydrogen already cause
high opacity, and photons are rapidly absorbed.

The EUV part of the ionizing stars in HII regions is poorly known;
it is inaccessible observationally and can only be known from stellar
atmosphere models. Ionizing stars in HII regions, mainly O and B stars and
sometimes Wolf-Rayet stars, possess strong winds, and strong non-ETL
effects are present in their atmospheres (Hillier \& Miller \cite{hillier};
Pauldrach \cite{pauldrach};
Schaerer \& Vacca \cite{schaerervacca}; Schaerer \& de Koter \cite{schaererdekoter};
Smith et al. \cite{smith}). In the EUV, various bound-free ionization
discontinuities are present, strongly affecting the spectra;
however, differences between
models by different authors can be significant (Morisset
\cite{morisset}).

The main constraint on these model spectra comes from the reconversion of
their energy into emission lines, which are more easily observed.
Morisset \cite{morisset},
for instance, used high ionization stages lines observable in the far
infrared to try to constrain the spectral energy distribution of
ionizing stars of
HII regions in the EUV. This kind of work requires the use of
photoionization codes, like CLOUDY (e.g. Ferland \cite{ferland}), NEBU
(e.g. Pequignot \cite{pequignot}),
PHOTO (e.g. Stasinska \cite{stasinska}).

However, more basic information can be
extracted from just the analysis of hydrogen recombination lines: in
the photoionization equilibrium, the number of ionizations
of neutral hydrogen is 
balanced
by the number of recombinations.
The number of ionizations can
easily be calculated from the (theoretical) ionizing spectrum, and the
number of recombinations can be counted in hydrogen recombination lines
that hardly depend on the plasma parameters (e.g. electronic
temperature).
The intensity of hydrogen recombination lines should thus simply
be proportional to the integrated luminosity of the ionizing spectra
in the EUV. This has been known since the pioneering works by Zanstra
\cite{zanstra} on planetary nebulae, but the measurements of EUV luminosities
by hydrogen Balmer recombination lines strongly rely on the hypothesis of the
optical thickness of the nebulae to Lyman continuum photons. In
planetary nebulae, Tylenda \& Stasinska \cite{tylenda} have shown
that the optical thickness strongly affects the measurements of
their EUV luminosities.\\

As far as  HII regions are concerned, 
direct measurements of the Lyman continuum for local starburst
galaxies show that either they are optically thick or only a low fraction
($\lesssim 10\%$) of ionizing photons are allowed to escape (Leitherer et al. \cite{leitherer_95}; 
Heckman et al. \cite{heckman}; Deharveng et al. \cite{deharveng}; 
Bergvall et al. \cite{bergvall}). 
On the other hand, also from direct Lyman continuum measurements, 
Steidel et al. \cite{steidel}  found  {\em no evidence} for the absorption of 
ionizing photons in high redshift (z$\simeq$3.4) Lyman Break Galaxies; they estimate a 
{\em lower limit} for their Lyman continuum escaping photons of 50\%.

 In the local universe, indirect measurements of lower limits of the escaping
fraction of ionizing photons are possible from the emission of the diffuse ionized 
gas (DIG): Zurita et al. \cite{zurita} estimate that the luminosity of the DIG in
local spiral galaxies can reach 25--70\% of their total ${\rm H\alpha}$ luminosity.
 One has to bear in mind that the actual 
escaping fraction can be higher, as some photons may leak sideways, perpendicularly to 
the galactic plane. In the Milky Way, Bland-Hawthorn \& Malloney  \cite{bland} 
estimate the upper bound to the
 leaking fraction of ionizing photons to be only of the order of 10\%, 
from the ionization of clumps in the magellanic stream.

 The understanding of the emission lines sequence in HII galaxies can 
bring some insight into this problem;
Stasinska et al. \cite{stasinskaetal} and
Stasinska \& Izotov \cite{stasinskaizotov} concluded from the statistical 
photoionization modelling of large samples 
(either published catalogues from spectrophotometric surveys or SDSS early data release
\footnote{Sloan Digital Sky Survey (www.sdss.org)}) 
that some systematic effect was present in
their optical thickness, e.g., that a fraction of HII galaxies was
optically thin, {\em and suggested that HII galaxies were becoming thinner with 
time.}
Castellanos et al. \cite{castellanos} reached similar
conclusions from the detailed photoionization analysis of 3 giant HII regions in nearby galaxies, e.g., that they were losing a significant  fraction of their
 ionizing photons (between 10-70 \%).

 Photoionization modelling, however, generally does not take into account the
 presence of dust, which can absorb Lyman continuum photons before they even have a chance 
to ionize the gas and then re-emits them in the  infrared. 
A significant fraction of the photons that are interpreted as escaping
photons may actually be used to heat up the dust. Assuming that no 
ionizing photon escapes, Inoue  \cite{inoue} infers from the comparison of radio 
(gas free-free emission) to infrared (dust emission) luminosities of individual HII regions
that the fraction $f_i$
of ionizing photons  used for photoionization varies between 0.4 and 0.7. 
Since he  assumes that {\em all} ionizing photons are used either to ionize the gas or heat up
the dust, the fraction $1-f_i$ used to heat up the dust should only be considered  as 
an upper limit, as some ionizing photons may simply escape. Nonetheless, this upper limit,
of the order of 0.6, is significant. 
Inoue et al. \cite{inoueetal} and Hirashita et al. \cite{hirashita} confirmed this result
 for spiral galaxies, finding  upper bounds for $1-f_i$ ranging from 0.4--0.7
for $1-f_i$.\\ 

It has to be emphasized that the problem of the optical thickness of the
ionized gas in HII galaxies is not restricted to these galaxies.
Reionization by the first hypermassive population III stars is a connected
question, as the gas ionization responds to the energy input and to the
incoming radiation in nearly the same way as in local HII galaxies.
Most likely, the gas constituting the intergalactic medium at this stage of
the evolution of the Universe is heterogeneous and clumpy.
Its interactions with primordial stars can only be understood in terms of
complex 3D hydrodynamical and photoionization modelling (Abel et al \cite{abel}),
whereas observational constraints are nearly non-existant. HII galaxies
might be the closest that we have to a
laboratory of the physical conditions of reionization.\\

Effects originating from the optical thickness, or geometrical effects, are
intrinsically hard to distinguish from effects arising from
uncertainties concerning the ionizing spectra  or from the presence of dust.
One way to examine the question is to look at the
metallicity of the ionizing clusters. The integrated spectra of an
ionizing population should
vary with metallicity, partly because of the variation of the metal opacities
of the stars that produce the ionizing spectra, but mostly because of the variations of
evolutionary tracks due to the metallicity (Schaller et al. \cite{schaller};
Charbonnel et al. \cite{charbonnel}; Schaerer et al. \cite{schaerera}, \cite{schaererb};
Mowlavi et al. \cite{mowlavi}; Girardi et al. \cite{girardi};
Fagotto et al. \cite{fagotto1}, \cite{fagotto2}; Bressan et al. \cite{bressan}).
A great part of the opacity in the atmospheres of hot stars in the EUV
is due to bound-free transitions of hydrogen and helium - and thus is not
affected by metallicity (see Leitherer et al. \cite{leitherer}).
 The amount of dust  is also expected to increase with
metallicity (Inoue et al \cite{inoueetal}).
Variations of the optical thickness with metallicity (which can easily be
measured from oxygen emission lines of the ionized gas)
should therefore be indicative of systematic effects
of the ionizing spectra,  or of the dust, and represent an important check on how to
segregate effects
of optical thickness from  other ones. HII galaxies are actually quite fit for this purpose, because
their HII regions are giant, with hundreds or thousands of ionizing
stars;
in this case, it is much easier to predict the ionizing radiation with
a population synthesis code than in normal HII regions, which only
possess a few ionizing stars, because the counting noise strongly affects the
predicted ionizing spectrum (Cervino \& Valls-Gabaud \cite{cervino}).\\

In this paper, we investigate effects of optical thickness in a sample of HII
galaxies by comparing the number of ionizing photons produced by stars and
the number of recombinations. In Sect. 2, we will describe the sample, and
our method to evaluate the number of ionizing photons from the stellar spectra.
In Sect. 3, we will describe how we can understand statistically the
evolution of the (mean) optical thickness with time, e.g., with the
nebular age, and how we can check for systematic effects of the model
ionizing spectra with metallicity. Finally, Sect. 4 will present the
results and a summary.

\section{Sample and methodology}

Our study is based on the spectrophotometric catalogue of HII galaxies
by Kehrig et al. \cite{kehrig}. This catalogue contains homogeneous
longslit observations of 188 spectra of individual
regions in a sample of 111 HII galaxies.
The choice of the regions was made by luminosity, the slit
being positioned along the brightest regions. These regions are typically
made up of bright knots, generally encompassing several
stellar superclusters, with sizes of a few hundreds of pc.

The Kehrig et al. catalogue includes most of the southern
hemisphere HII galaxies listed in Terlevich et al. \cite{t91}, but with a
higher signal/noise.
It also includes nearby HII galaxies that have been discovered since the
publication of Terlevich et al.'s catalogue. The stellar population content
of the galaxies from this catalogue, as derived from their
spectra, has already been discussed in Westera et al.
\cite{westera_04}.

Westera et al. \cite{westera_04} based their study on the analysis of
modified Lick
indices (Worthey \cite{worthey}; Worthey et al. \cite{worthey_2}).
What we want to do here is to use a similar population synthesis method,
based
on a full spectra fitting in the 4000-7000~${\rm \AA}$ range,
with the purpose of predicting the ionizing spectrum in the EUV, for
$\lambda \leq 912 \AA$.

A best-fit procedure was employed using three different libraries of
simple stellar populations (SSPs).
The first SSP library (hereafter the ``BC99'' library) was
produced using the Bruzual and Charlot 2000 Galaxy Isochrone
Spectral Synthesis Evolution Library (GISSEL) code
(Charlot \& Bruzual \cite{charlot_91}; Bruzual \& Charlot \cite{bruzual_93}, \cite{bruzual_00}), implementing the Padova
2000 isochrones (Girardi et al. \cite{girardi_00}) combined with the BaSeL 3.1
``Padova 2000'' stellar library (Westera et al. \cite{paperiii};
Westera \cite{diss}).
The second SSP library, ``Starburst'',
consists of spectra from the STARBURST99 data package
(Leitherer et al. \cite{leitherer}), using the option of including nebular
continuum emission (Fig. 1 on the STARBURST99 web site). It implements the
BaSeL 2.2 library (Lejeune et al. \cite{lejeune_97}, \cite{lejeune_98}), and for stars with strong mass loss, it also takes
into account extended model atmospheres by
Schmutz et al. \cite{schmutz},
combined with the Geneva isochrones (Meynet et al. \cite{meynet};
Schaller et al. \cite{schaller}; Schaerer et al. \cite{schaerera}, \cite{schaererb}; Charbonnel et al. \cite{charbonnel}).
For old populations, the ``BC99'' spectrum was used, since the Starburst99
data package only contains spectra up to 900 Myr.
In addition to these two libraries - which were already used in
Westera et al. \cite{westera_04} - we also used a library with higher
spectral
resolution, ``BC03'', produced by employing the 2003 version of the
GISSEL code (Bruzual \& Charlot \cite{bruzual_03}) and the Padova 1995
isochrones (Fagotto et al. \cite{fagotto2}; Girardi et al. \cite{girardi})
combined with the STELIB
(Le Borgne et al. \cite{leborgne}) stellar library.
As the nebular continuum emission is important for the spectral shape
of young populations, it was also added to the spectra in the ``BC99''
and ``BC03'' libraries, in the  way described by
Leitherer et al. \cite{leitherer}. \\
Table~\ref{libraries} provides a summary of the evolutionary tracks and
stellar libraries used for the three SSP libraries.
   \begin{table}
   \begin{center}
      \caption{Evolutionary tracks and stellar libraries used in the SSP libraries.}
      \label{libraries}
         \begin{tabular}{lll}
            \hline
 SSP library   & Tracks & Stellar library \\
            \hline
 ``BC99''      & Padova 2000 & BaSeL 3.1 ``Padova 2000'' \\
 ``Starburst'' & Geneva      & BaSeL 2.2 \\
 ``BC03''      & Padova 1995 & STELIB \\
            \hline
         \end{tabular}
   \end{center}
   \end{table}

Since for the present purpose, it is more important to reproduce the
shapes of the galaxy spectra than to find the actual parameters of the
sub-populations, we fit the full spectra from 3910 \AA\ to
6880 \AA\ instead of single spectral indices, except for those various
parts of the spectra showing ``contaminations'' from
different emission and/or telluric line sources.
Table~\ref{cutouts} lists all the regions that were not used for the fits.
   \begin{table}
   \begin{center}
      \caption{Wavelength ranges that were not used for the spectral fit.}
      \label{cutouts}
         \begin{tabular}{ll}
            \hline
 Range (\AA) & ``Contamination'' source \\
            \hline
 3960-3975 & H$\epsilon$+[NeIII]3967 \\
 4095.25-4106.5 & H$\delta$ \\
 4331-4370 & H$\gamma$+[OIII]4363 \\
 4465-4479 & HeI 4471 \\
 4848-4871 & H$\beta$ \\
 4943-5020 & [OIII]4959+5007 \\
 5446-5574 & telluric lines \\
 5867-6320 & HeI 5876, [OI]6300, [SIII]6312 \\
 6375-6470 & telluric lines \\
 6520-6600 & H$\alpha$+[NII] \\
 6700-6740 & [SII] 6717, 6731 \\
            \hline
         \end{tabular}
   \end{center}
   \end{table}

In the present paper, we modelled the actual
population as being composed of an old, an intermediate,
and a young stellar population (in Westera et al. \cite{westera_04},
we only decomposed the actual population as a young+intermediate one
and an old one).
The characteristics and free parameters of the three populations are
summarised in Table~\ref{parameters}.

     The spectra were first blueshifted to their rest wavelength. 
    Internal gas extinction was then corrected for, using the observed 
   ${\rm H_{\alpha}/H_{\beta}}$ emission lines ratio. The resolution of our 
    spectra was sufficient ($\simeq 5\AA$) to easily deconvolve ${\rm H_{\alpha}}$ from 
    the [NII] doublet at ${\lambda\lambda 6548,6583 \AA}$ (see Kehrig et al. \cite{kehrig}).  As explained in 
    next section, we did not adopt the emission line intensities from Kehrig et al. \cite{kehrig}, 
    but we remeasured them  after fitting the stellar continuum by
    population synthesis. In HII galaxies spectra, ${\rm H_{\alpha}}$ and ${\rm H_{\beta}}$
    are present both in absorption (from the stars) and in emission (from the ionized gas);
    our procedure allowed us to disentangle both components. 
     The repartition of  gas and stars  inside a galaxy lets them be affected in different
    manners by extinction (see Calzetti et al. \cite{calzetti}).
     We dereddened the stellar continuum spectra employing the formulas
   $E(B~-~V)=0.665\times C_{H\beta}$ (Sampson \cite{leda}) and
   $E(B~-~V)_{cont}=0.44\times E(B~-~V)_{gas}$
   (thus, $E(B~-~V)_{cont}=0.2926\times C_{H\beta}$) and the extinction law of
   Fluks \cite{fluks}. The factor $0.44$ was introduced to correct
   for systematic differences between the extinction as derived from the stars and
   from the ionized gas (Calzetti et al. \cite{calzetti}).


The best-fitting population was found by a $\chi^2$ algorithm.
To be able to calculate the $\chi^2$ estimator, the empirical
and theoretical spectra have to be on the same wavelength grid (at least in
the range used for the fit).
This was done by rebinning the empirical spectra to the resolution of the
theoretical spectra using a Gaussian kernel function with a full width at
half maximum (FWHM), corresponding to the resolution of the theoretical
spectra (20 \AA\ for ``BC99'' and ``Starburst''  and 1 \AA\ for ``BC03''). \\
In the end, a selection of acceptable solutions was made. All spectra
that had too bad a signal-to-noise or a strange shape,
indicating calibration problems in the data reduction,
were eliminated from the sample. This selection was done by eye.
It left us with solutions for 105 spectra of individual regions in
72 galaxies.
%
%
%

\begin{figure*}
\centering
\includegraphics{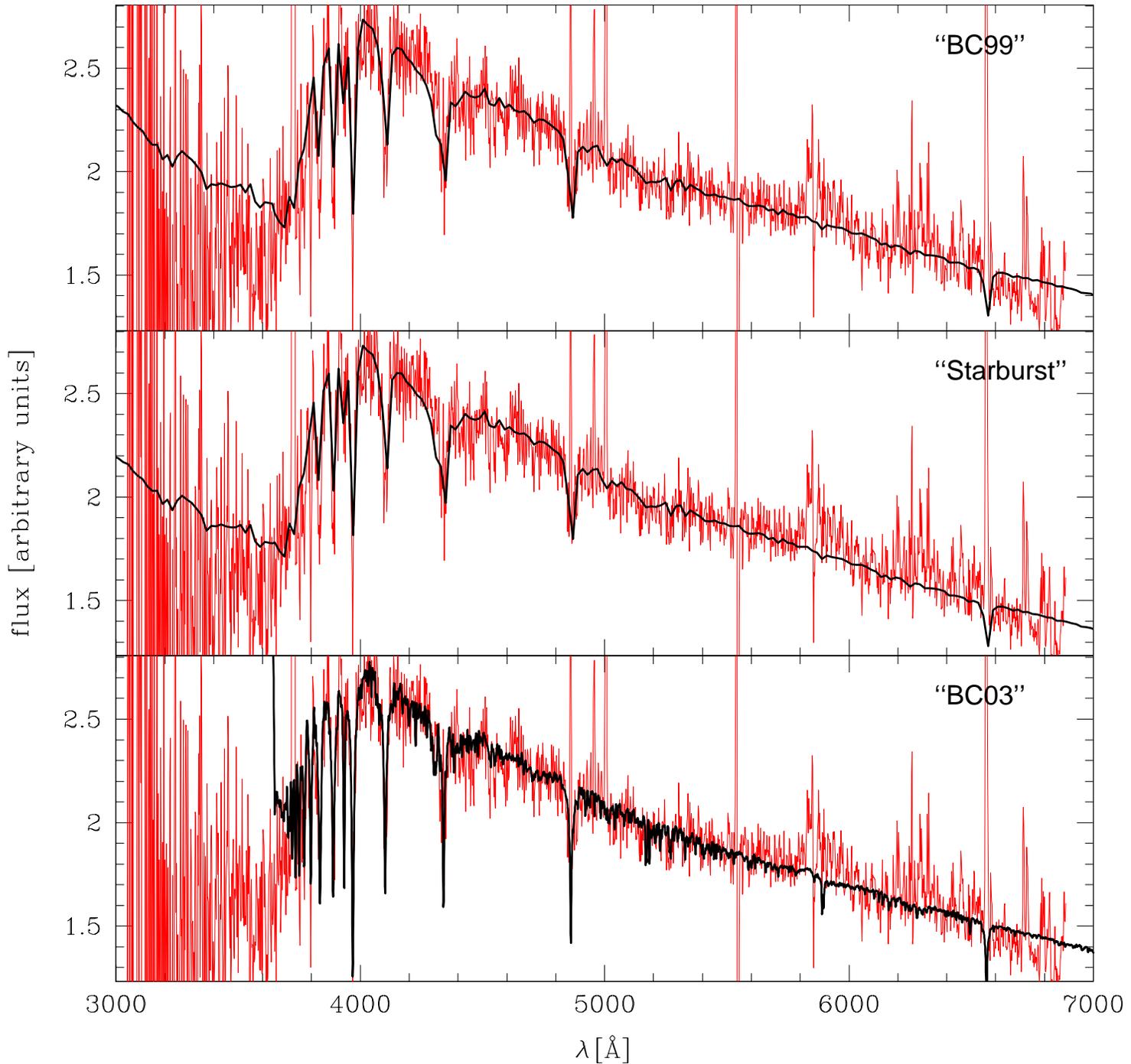}
\caption{Example of a ``best fit''. The thin (red) lines represent
       the empirical spectrum (UM 137(W), taken on August, 18,
       1998); the thicker black lines show the best-fitting spectra
       obtained using the different SSP libraries.}
\label{spec}
\end{figure*}

\begin{figure*}
\centering
\includegraphics{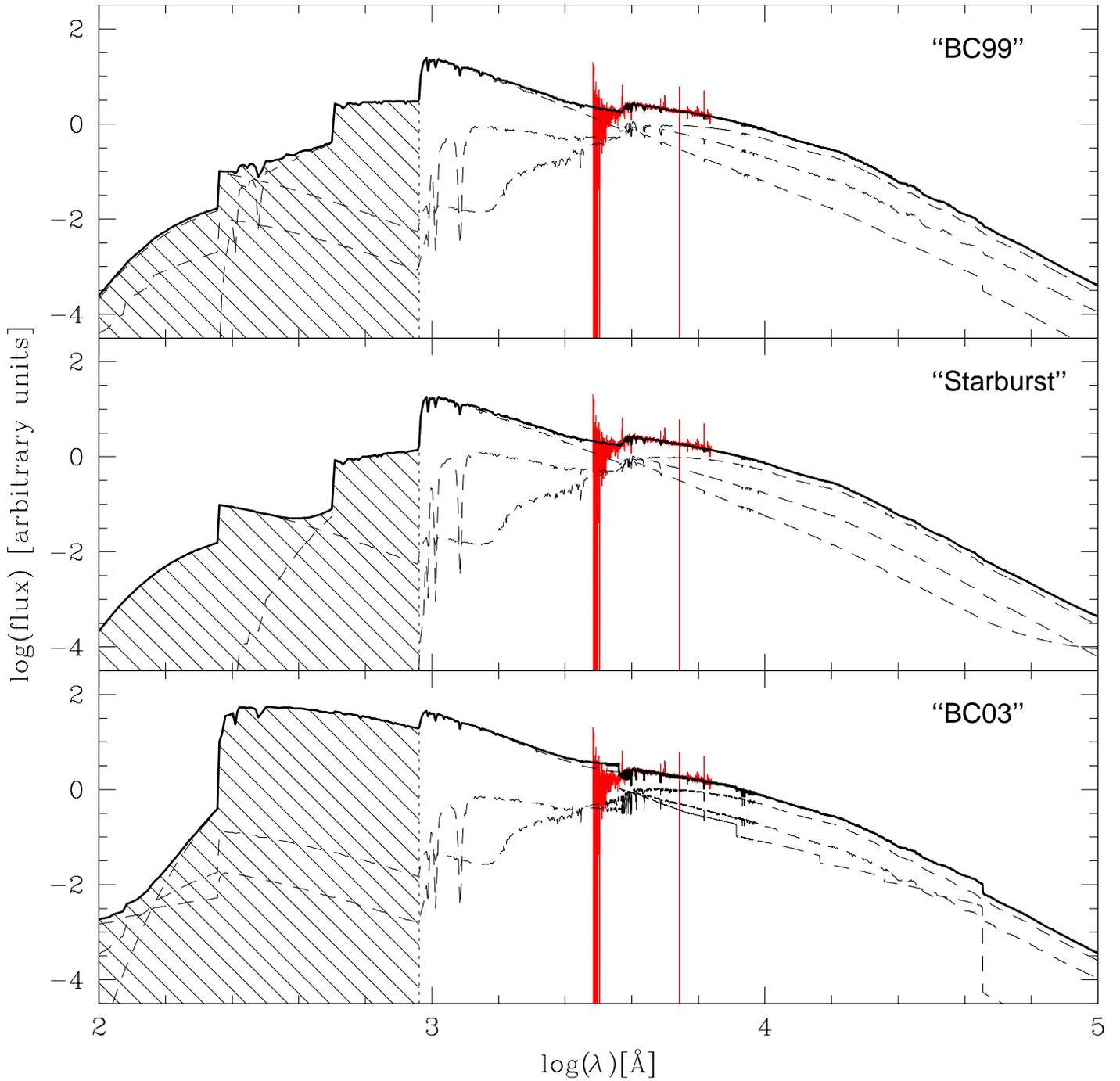}
\caption{Synthesised spectra over the whole wavelength range.
       The thin (red) lines represent the empirical spectrum
       (as in Fig.~\ref{spec}),
       the solid lines show the best-fitting spectra, and the
       dashed lines show the decompositions of the best-fitting
       spectra into young, intermediate, and old
       populations, respectively.
       The shaded regions represent the radiation shortward of
       the Lyman limit (912 \AA), which is responsible for the
       ionization of the surrounding gas.
       Note that both axes are on a logarithmical scale, in contrast
       to Fig.~\ref{spec}.}
\label{spec2}
\end{figure*}

The best-fitting population parameters for the fits with the ``BC03''
library can be found in columns 4 to 8 of Table~\ref{EWparameters}.
Figure~\ref{spec} shows an example of an observed spectrum and the
resulting composite spectra in the wavelength range where the fits
were made, and Fig.~\ref{spec2} represents the extrapolations
to the UV and EUV. \\
The introduction of a third partial population again serves the purpose of
better reproducing the spectral shapes.
In reality, the quality of our spectra is not high enough to determine
the parameters (metallicities, ages, and masses) of three sub-populations,
so the parameter values given in Table~\ref{EWparameters} should be
taken with a grain of salt.
However, the parameters that can directly be compared with those
derived in Westera et al. \cite{westera_04} agree reasonably well:
We obtain the same value for the parameter $(M_{y}+M_{i})$:$M_{o}$ as
for $(M_{y+i})$:$M_{o}$ in Westera et al. \cite{westera_04} for 38\% or 49\%
of the spectra (depending on the spectral library, see next paragraph),
and in another 36\% or 28\%, these two parameters differ only by one ``step''
in the grid.
In 49\% or 46\% the luminosity-weighted (geometrical) mean of $age_{y}$ and
$age_{i}$ differs by less than two ``steps'' from the $age_{y+i}$
calculated in Westera et al. \cite{westera_04}, which corresponds to a
factor of less than 5 in $age_{y+i}$.
%
%
%
%
%
   \begin{table}
   \begin{center}
      \caption{Possible values of the population parameters.}
      \label{parameters}
         \begin{tabular}{ll}
            \hline
 Parameter & Possible values \\
            \hline
 $(M_{y}+M_{i})$:$M_{o}$    & 0:1, 1:100, 1:30, 1:10, 1:3, 1:1, 3:1, 10:1, 1:0 \\
 $M_{y}$:$M_{i}$    & 0:1, 1:30, 1:10, 1:3, 1:1, 3:1, 10:1, 30:1, 1:0 \\
 $age_{y}$          & 1, 2, 3, 4, 5, 6, 7, 8, 9 Myr \\
 $age_{i}$          & 10, 20, 50, 100, 200, 500 Myr \\
 $age_{o}$          & fixed at 5 Gyr \\
 ${\rm [Fe/H]}_{y}={\rm [Fe/H]}_{i}$ & taken from Cuisinier et al. \cite{cuisinier} or -1 \\
 ${\rm [Fe/H]}_{o}$ & fixed at -1.5 \\
            \hline
         \end{tabular}
   \end{center}
   \end{table}

In Westera et al. \cite{westera_04}, we discussed the differences arising from
adopting the Starburst 99 (Leitherer et al. \cite{leitherer}) or the BC93
(Bruzual \& Charlot \cite{bruzual_93}) codes for the young population (the old
one being modelled with the BC93 code in both cases).
In this paper we enhance these results by modelling both the
old and young populations with the BC03 (Bruzual \& Charlot \cite{bruzual_03})
version of their code.
Each of these
codes, or code versions, has their advantages and inconveniences:
Starburst 99 is particularly fit for the study of young populations
in the visible/ultraviolet, but does not extend to ages older than 1Gyr; BC93
is made to model all kinds of populations, but is  optimized
for old ones. Finally, the BC03 version of the code
(Bruzual \& Charlot \cite{bruzual_03}) includes
empirical spectral libraries of higher resolution, BC-2000
(Pickles \cite{pickles}; Bruzual \& Charlot \cite{bruzual_03}) and STELIB
(le Borgne et al. \cite{leborgne}), and, therefore, is better suited
for calculating Lick
indices (because of its higher spectral resolution) than Starburst 99 or BC93.
However, it has a poor coverage of parameter space for the ionizing hot
stars. Furthermore, the extension of the BC03 version
to wide and comprehensive
wavelength ranges is made by combining STELIB spectra in the visible,
Pickles library (\cite{pickles}) in the UV and IR, and BaSeL theoretical
spectra beyond 2.5~$\mu$m of for wavelengths shorter than 1200~${\rm \AA}$.
Comparing the results of different codes or
versions, we hope to be able to better assess possible systematic effects. \\

\section{Evaluation of the photon loss rate}

\begin{figure}
\centering
\includegraphics[width=9cm]{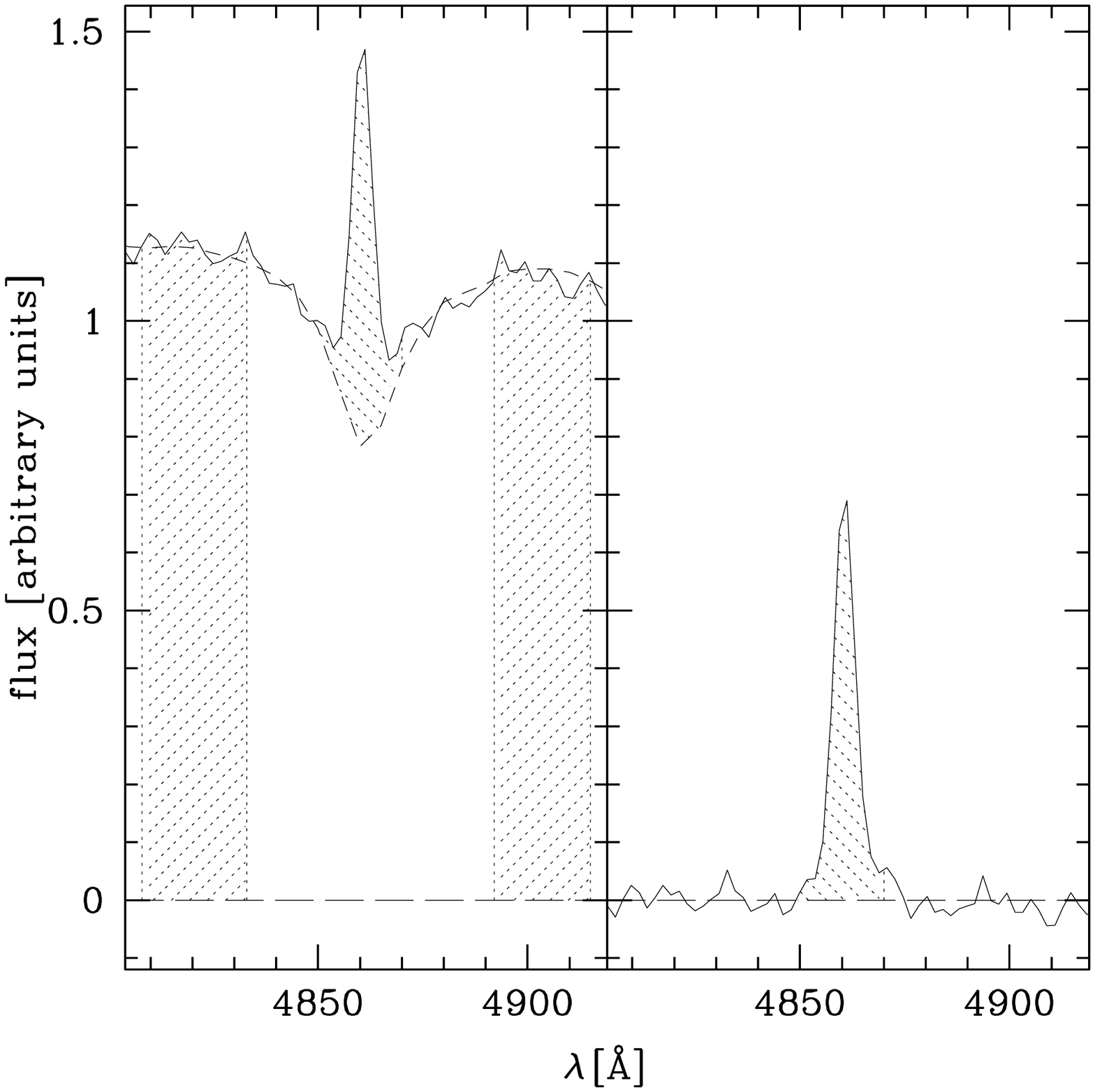}
\caption{Illustration of how the H$\beta$ emission line equivalent width was
       measured through the example of UM137(CentE).
       In the left panel, the solid line represents the empirical spectrum,
       whereas the short-dashed line shows the best fit using the
       BC-2000 stellar library.
       The shaded region between these two lines (shaded with lines running
       from the upper left to the lower right) shows the area used to
       calculate the emission line strength, and the shaded region
       (lines running from the upper right to the lower left) between
       the solid line and zero level (the long-dashed line) shows the
       location of the continuum band.
       The right panel shows the emission line after subtracting the
       (rebinned) best fit. The shaded area corresponds to the shaded
       area between the spectrum and the best fit from the left panel.
       Here, too, the zero level is shown as a long-dashed line to
       show the quality of the subtraction.}
    \label{Hbeta}
\end{figure}

We evaluate the number of ionizing photons with the method exposed in the
previous section from an extrapolation of the
ionizing spectrum to the EUV, based on population synthesis.
The number of recombinations
can also be calculated from a recombination line of hydrogen, like ${\rm H\beta}$.
If the number of predicted ionizing photons is correct, in the case of an
optically thick ionized nebula, both numbers are equal (in the photoionization
equilibrium).
In an optically thin nebula, however, not all the ionizing photons will
actually be able to ionize the gas, and some will escape. The ratio of the
number of (predicted) ionizing photons to the number of recombinations
can thus be considered an indicator of optical thickness. We define:
   \begin{equation}
   {\rm \Delta \log Q(H^0) = \log Q(H^0)_{ion} - \log Q(H^0)_{rec} ,}
   \end{equation}
where ${\rm Q(H^0)_{ion}}$ is the number of 
ionizing photons, as determined from
the extrapolation of the ionizing spectrum to the EUV, based on population
synthesis, and ${\rm Q(H^0)_{rec}}$ is the number of recombinations as calculated from 
the ${\rm H\beta}$ recombination line,  after correction from absorption, as estimated
from ${\rm C_{H\beta}}$. \\
To properly measure the equivalent width of the H$\beta$
emission line, we had to correct its strength for the reduction it suffers
from the underlying absorption line.
For this purpose, we first made a best fit of the continuum  in the 3910-6880~${\rm \AA}$
wavelength range, 
in a similar way to Westera at al. \cite{westera_04},
but putting extra weight on the H$\delta$ index (the wings of the H$\delta$
absorption line) and using the theoretical spectra from the BC-2000
stellar library.
Although this library only exists for solar metallicity, its resolution is
much higher than for the other libraries, and the best-fitting spectra
reproduce the shape of the absorption lines better than the fits
obtained with the other libraries.
Note that these fits were only made for calculating the ${\rm H\beta}$
equivalent widths and have nothing to do with the population synthesis
described in Sect. 2.
After subtracting the (rebinned) best-fitting spectra from the galaxy
spectra, we could measure the strength of the emission line and thus
calculate the equivalent widths EW(H$_{\beta}$).
This is illustrated in Fig.~\ref{Hbeta}.
Results are given in column 2 of Table~\ref{EWparameters}. \\
Using these definitions, ${\rm \Delta \log Q(H^0)}$ will be equal to 0 for
optically thick nebulae and increase for nebulae getting optically
thinner and thinner.
The ${\rm \Delta \log Q(H^0)}$ values of individual spectra can be
found in column 3 of Table~\ref{EWparameters}.

As we are more interested here in giant HII regions than in the HII galaxies
themselves, we will focus our discussion on the spectra of the individual
regions from Kehrig et al.'s \cite{kehrig} catalogue, and not on the spectra
of the individual galaxies. As a matter of fact, we checked if our
results were the same considering only the integrated spectra of galaxies, and we found
excellent agreement.

We could have some systematical bias due to the fact that we observed
the galaxies, or giant HII regions, through apertures defined by the slit,
and thus not encompassing the  HII regions (galaxies) entirely. As we centred
the slit on the brightest regions, we estimate that the quantity
of nebular
H$\beta$ photons we lose should not exceed a factor of $\simeq 3$
(0.5 in logarithm). \\

\subsection{Evaluation of the ionizing spectra}

It is intrinsically hard to disentangle effects originating
from the ionizing spectra from effects arising from the optical thickness 
of the nebulae: For a given HII region, an erroneous modelling of the
number of ionizing photons can give  results equivalent to the ones that would
be attributed to an optically thin nebula. 
The accuracy of the modelling of the stellar spectra in the EUV can 
be assessed by the comparison of line intensities of ions of different
ionization potentials, as in Morisset et al. \cite{morisset}.
Much more basic information can, however, be extracted from systematical  tendencies
of ${\rm \Delta \log Q(H^0)}$ with metallicity.
If the predicted ionizing
spectra are wrong for some reason, their systematical excess or deficiency
in ionizing photons will vary with metallicity because of the 
intrinsic change of the integrated ionizing spectra  (see 
Leitherer et al. \cite{leitherer}) and will be seen in
${\rm \Delta \log Q(H^0)}$, if plotted as a function of metallicity.
On the other hand, geometrical effects (optical thickness) are not
expected to vary with metallicity and will thus just produce a systematical
offset in ${\rm \Delta \log Q(H^0)}$. 

\begin{figure}
\centering
\includegraphics[width=9cm]{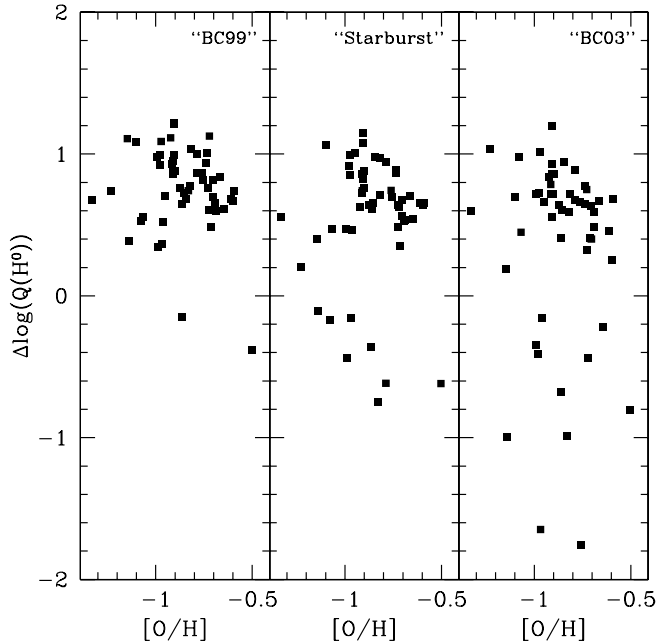}
\caption{Logarithmic ratio of the predicted number of ionizing photons to
       the number of recombinations as a function of the oxygen abundance
       from the gas.}
\label{dqoOH}
\end{figure}

Figure~\ref{dqoOH} shows a comparison of
${\rm \Delta \log Q(H^0)}$
with the oxygen abundance as derived from emission lines of the gas
(Cuisinier et al. \cite{cuisinier}). Most giant HII regions have
${\rm \Delta \log Q(H^0)} > 0$, indicating that they let ionizing
photons escape, since they are optically thin. A huge scatter exists, however, with some giant
HII regions reaching
negative values of ${\rm \Delta \log Q(H^0)}$. Nonetheless, as we will show in
Sect. 3.3, we believe it to be due to uncertainties
in the population synthesis fitting
procedure. If we look at the bulk of the giant HII regions, however,
they lie at a constant value of ${\rm \Delta \log Q(H^0)}$, around 0.8, with
no clear tendency for oxygen abundance. 

 Another point that could affect our evaluation of the number of
ionizing photons is  dust absorption. As stated in Sect. 2, we took
great care in correcting the observed spectra for absorption before 
performing any fitting procedure. We did this by measuring the reddening
from the  observed ${\rm H\alpha/H\beta}$ line ratio. Absorption for the 
stellar population, however, is not necessarily the same as for the 
gas phase: the geometrical repartition is not identical in respect to the
dust. We applied a correction factor on the reddening for the stellar population 
to compensate for this effect
(Calzetti et al. \cite{calzetti}), but it is important to check that no 
remaining differential effect exists. Furthermore, dust can absorb high energy ionizing
photons, affecting our evaluation of ${\rm \Delta \log Q(H^0)}$.

  The dust quantity should 
increase  with metallicity; any effect due to an erroneous 
evaluation of the absorption should affect the variation of   
${\rm \Delta \log Q(H^0)}$ with
 metallicity, but, as can be seen in  Fig. ~\ref{dqoOH}, no clear tendency 
can be detected.

 This might seem contradictory to the results by Inoue et al. \cite{inoueetal}:
From the analysis of the infrared emission of HII regions, due to the dust,  and their radio 
emission, due to the gas, they find that the fraction of Lyman continuum  photons
absorbed by the dust increases with metallicity (assuming that no photons are leaking).
The metallicity range of their HII regions is different, however, and extends 
to higher metallicities than our HII galaxies, though some overlap exists.
A closer inspection of their data shows that at low metallicities, equivalent to the ones 
we encounter in our sample ([0/H]$\leq-0.5$), the fraction of Lyman continuum photons used for dust heating
$1-f_i$ is rather constant at $\simeq 0.2$. It only increases for higher metallicities.
One must bear in mind that, as stated in the introduction, since they do not consider the possibility
of photons leaking out, $1-f_i$ is an upper limit to the fraction of Lyman continuum photons used
to heat up the dust. 
  

 Thus, from our analysis of the variation of
${\rm \Delta \log Q(H^0)}$ with metallicity, (i) we cannot put 
into evidence any deficiency from the theoretical ionizing spectra, and (ii)
we cannot rule out effects of absorption by the dust, if they
are constant with metallicity. Inoue et al. \cite{inoueetal} showed 
that such an effect  exists, but at the metallicities of our 
HII galaxies sample, it should be lower than 20\%.

\begin{figure}
\centering
\includegraphics[width=9cm]{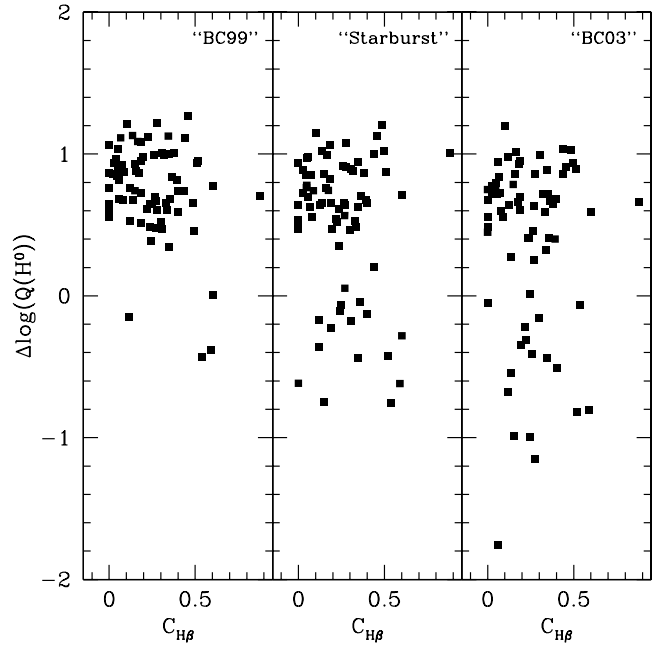}
\caption{Logarithmic ratio of the predicted number of ionizing photons to
       the number of recombinations as a function of the gas reddening
       parameter ${\rm C_{H\beta}}$.}
\label{dqoCHB}
\end{figure}

As a consistency check, in  Fig.~\ref{dqoCHB} we compare  
${\rm \Delta \log Q(H^0)}$ with the measured ${\rm C_{H\beta}}$
values;  effects due to the dust,  provoking either
an erroneous evaluation of the absorption for the stellar population in the
visible
or  an absorption of high energy ionizing photons, should 
vary with the dust quantity, e.g., the reddening. As can be seen
in Fig.~\ref{dqoCHB}, no tendency can be detected.



\subsection{Temporal evolution of the optical thickness to the ionizing
photons}
\label{evolution}

The equivalent width in ${\rm H\beta}$ is a quantity that 
monotonically
decreases with time within a given HII region, or even within a given star
formation episode of an HII galaxy. The number of ionizing photons decreases
with time, decreasing the ${\rm H\beta}$ intensity, and the continuum at
the same wavelength rises.
The equivalent width in ${\rm H\beta}$ has been used extensively
 as a nebular age indicator by Stasinska et al. \cite{stasinskaetal},
Stasinska \& Izotov \cite{stasinskaizotov}, Terlevich et al. \cite{terlevich},
and Dottori \& Bica \cite{dottoribica}, as was first suggested by
Dottori \cite{dottori}.\\

The underlying non-ionizing population, however,  blurs  this
picture by raising the continuum; the ${\rm H\beta}$ equivalent width anti-correlates
with oxygen abundance (though with a huge scatter) (Terlevich et al. \cite{terlevich}),
 definitely pointing towards a contamination of the continuum by previous populations. 
Furthermore, stellar population studies (e.g., Westera et al.
\cite{westera_04}; Kong et al. \cite{kong}; Cid-Fernandes et al. \cite{cidfernandes}) 
indicate that stellar formation events last for times of the order of at least 100 Myr, not
10 Myr (which is the photoionization timescale). A stellar formation event in an HII galaxy {\em
naturally} accumulates an intermediate age population (between 10 and a few 100 Myr) that 
participates in the continuum budget. 

The two effects (i.e. (i) the accumulation of old populations from previous bursts, and (ii) the accumulation
of an intermediate age population due to the present burst) have  opposite effects in terms
of interpreting ${\rm H\beta}$ equivalent widths as star-forming event age indicators. 
The first one 
only introduces scatter in the relation, while the second one  helps to decrease  ${\rm H\beta}$
equivalent widths with increasing burst ages. This latter  effect happens, however,  in a manner 
that may not be as simple to evaluate  
as can  be naively done by simply comparing with photoionization 
by single aged stellar populations.

\begin{figure}
\centering
\includegraphics[width=9cm]{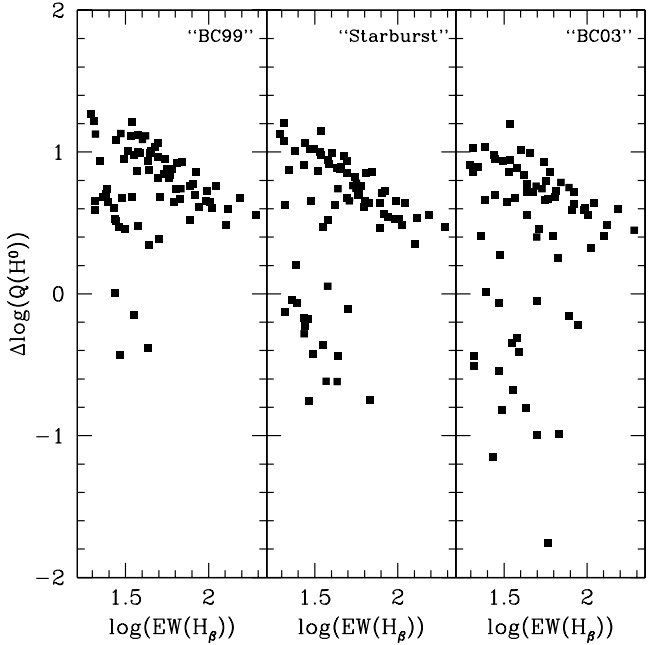}
\caption{Logarithmic ratio of the predicted number of ionizing photons to
       the number of recombinations as a function of the equivalent width
       of ${\rm H\beta}$.}
\label{dqoEWhb}
\end{figure}

Figure~\ref{dqoEWhb} shows a plot of ${\rm \Delta \log Q(H^0)}$
versus the equivalent
width in ${\rm H\beta}$, 
as calculated in the previous
subsection. As in Fig.~\ref{dqoOH}, we can see a
clustering of most data in ${\rm \Delta \log Q(H^0)}$ around a positive value
of 0.8. A big scatter exists as well, and as we will explain in
Sect. 3.3., we believe that we can explain it just from the uncertainties in the
population synthesis process. However, in contrast to
Fig.~\ref{dqoOH}, we see
a strong correlation of the clustered data with the equivalent width in
${\rm H\beta}$. ${\rm \Delta \log Q(H^0)}$, which can be considered an
indicator of the number of ionizing photons escaping the nebulae, increases with
the decreasing equivalent width in ${\rm H\beta}$, i.e., time.\\

This result may seem counter-intuitive, as the quantity of matter
absorbing the ionizing photons remains the same and the number of ionizing
photons decreases with time: the optical thickness should increase with time.
However, if we think of HII regions not as homogeneous media, but
consider them in their complexity, with all their clumps, filaments, and
substructures, we can understand the phenomenon, if we assume that
the nebulae are optically thin in some directions, and thick in others.
This can be described by a covering factor $f_c$, characterizing
the fraction of solid angle within which the nebulae are optically thick.
The increase of the number of ionizing photons escaping the nebulae with time
can thus be understood if the covering factor $f_c$ decreases with time.
This has already been
pointed out by Stasinska et al. \cite{stasinskaetal} and Stasinska \& Izotov
\cite{stasinskaizotov}, based on other
arguments (the time sequence of the intensities of collisional lines
in HII galaxies). The decrease of the
covering factor $f_c$ with time can be understood if the structure of HII
regions always becomes more complex with time, continuing to subfragment
and thus to form more and more ``holes'' in their structure while they expand,
always letting  more ionizing photons escape.
Tenorio-Tagle et al. \cite{tenorio} reach similar conclusions from the 
theoretical modelling of the hydrodynamical evolution of giant HII regions
under the effect of both photoionization and stellar winds in a clumpy
 interstellar medium.\\

Of course, the  absorption of ionizing photons by dust would provide an alternative
explanation; it would, however, require a mechanism that  increases the absorption
of ionizing photons by dust with time. As such a mechanism seems  
rather artificial to us, we prefer to stick to the leaking photons explanation.

The biggest problem with leaking photons is that they are not confirmed by the 
direct observation of the Lyman continuum in local starburst galaxies, at least in 
the quantities we predict (Leitherer et al. 
\cite{leitherer_95}; Heckman et al. \cite{heckman}; Deharveng et al. \cite{deharveng};
Bergvall et al. \cite{bergvall}), although is not the case at higher redshifts (Steidel
 \cite{steidel}). The total number of  galaxies  that have been observed in their Lyman continuum
is, however, quite small, just a little over  10. These galaxies
have been selected on the basis of the intensity of their starbursts. In the explanation
we propose, the youngest starbursts, i.e., ones with  the highest ${\rm H\beta}$ equivalent
widths, which would be interpreted as the ones having the most intense starbursts, 
have the lowest fraction of escaping ionizing photons. We evaluate this fraction
in these galaxies to be $\simeq$70\%, i.e., ${\rm \Delta Q(H^0)}\simeq 3$ on a linear 
scale; one has to take into account that our evaluations of ${\rm \Delta Q(H^0)}$ 
are certainly systematically over-evaluated by aperture effects by a factor that can 
be as high as  $\simeq 3$ (see Sect. 2). Adding some systematic absorption
by dust (which we, however, do not believe can be much higher than $\simeq 20\%$), we can easily
 lower the actual escaping fraction of ionizing photons in those galaxies to $\simeq 10\%$,
which is what is believed to be the upper limit for escaping photons from Lyman 
continuum observations. This is, however, certainly {\em not} the case for the most evolved, 
lowest ${\rm H\beta}$ equivalent width galaxies in our sample, where ${\rm \Delta Q(H^0)}$
reaches values higher than 10.

\subsection{Accuracy of the population synthesis modelling}

The interpretation of the data presented above highly relies on our ability
to evaluate the spectrum in the EUV from a population synthesis over the
observed range between 4000-7000 ${\rm \AA}$.
Population synthesis is an ill-conditioned problem, subject to all kinds
of degeneracies, and it is important to test its ability to recover what we
want.\\

\subsubsection{Comparison of the different libraries}

We first compared the results given by three different stellar population codes (or combinations),
 ``BC03'', ``BC99'', and ``Starburst''. The results commented on in Sect.~\ref{evolution} are overall consistent
with the three different codes, but some differences do exist. It is noteworthy that, as far
as the correlations with the ${\rm H\beta}$ equivalent widths are concerned, the best results are
obtained with ``Starburst'' (Fig.~\ref{dqoEWhb}). ``BC99'' gives nearly as 
good correlations, and with fewer outliers. Finally, results obtained with ``BC03'' do not give
any improvement; on the contrary, the correlation presented, though definitely existing, is not as 
strong as that with the two other libraries, and the number of outliers is higher.

Though this cannot be considered  definite evidence,
since effects on the number of ionizing photons is what 
we want to assess, this could be indicative of deficiencies of the BC03 stellar spectra 
library in the EUV. The BC03 library is made up of the extension of the empirical STELIB 
library in the visible (Le Borgne et al. \cite{leborgne}) to other wavelengths, with empirical spectra when  
available, and using theoretical spectra when empirical spectra were not available. The connection of the different spectra
of different origins has been based on evaluated physical parameters, or rather, parameter classes, 
introducing some inconsistency in the process, which is not present in entirely theoretical libraries; one must
 remember that the EUV part of the spectra is theoretical in {\em all} libraries.\\ 

\subsubsection {Monte-Carlo simulations}

As a matter of fact,
in our present problem, we are not so much interested in the underlying
population parameters, as in our ability to recover a given spectrum
from observations over a limited wavelength range.\\

To do this, we took the synthesised spectra over the whole
91 \AA\ to 160 $\mu$m, corresponding to the population mixtures derived
from the population synthesis over the observed range
(4000-7000 $\rm \AA$).
We added  Gaussian noise, such that the S/N ratio
varied from 2 to 40,
corresponding to the observed S/N ratio distribution
(see Kehrig et al. \cite{kehrig}), to the fitted synthetic spectrum (assumed to be our true prior), and recovered new
synthesised spectra through the same population synthesis process over the
4000-7000 ${\rm \AA}$ range. We assumed the first spectra to give the
true number of ionizing photons, giving the number of recombinations counted
in ${\rm H\beta}$ under the hypothesis of the nebulae's optical
thickness.
From the latter spectra, we compute the number of ionizing photons by
the population synthesis process, and we obtain ${\rm \Delta \log Q(H^0)}$
from the logarithmic difference of both numbers.
To obtain reliable statistics, we performed 1620 such Monte-Carlo
simulations for each of the three SSP libraries.

\begin{figure}
\centering
\includegraphics[width=9cm]{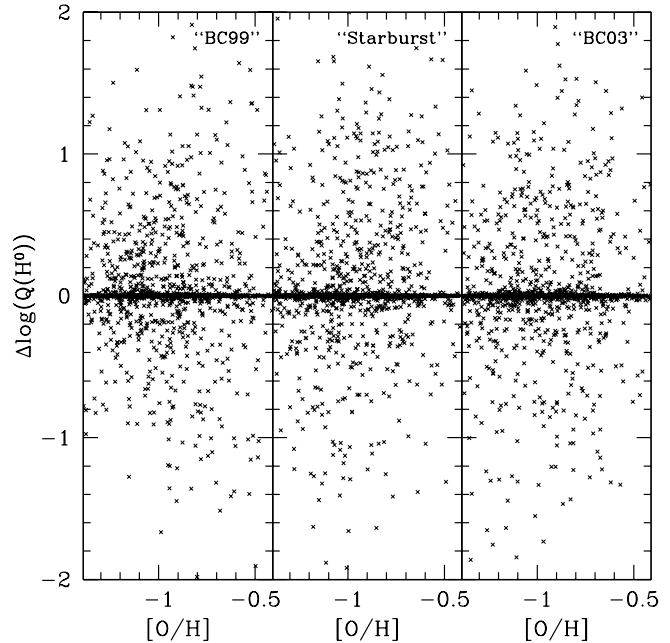}
\caption{Monte Carlo simulation of the logarithmic ratio of
       the predicted number of ionizing photons to
       the number of recombinations as a function of the oxygen abundance
       from the gas for optically thick nebulae.}
    \label{dqoOHmc}
\end{figure}

\begin{figure}
\centering
\includegraphics[width=9cm]{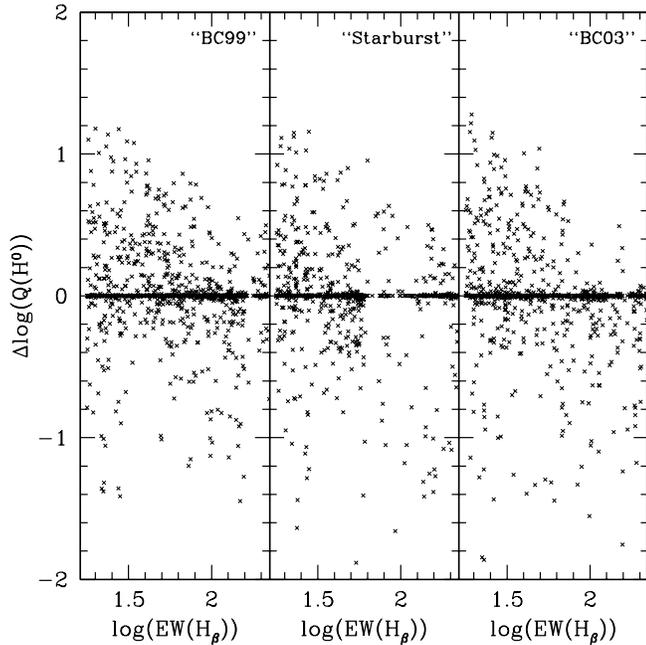}
\caption{Monte Carlo simulation of the logarithmic ratio of
       the predicted number of ionizing photons to
       the number of recombinations as a function of the equivalent width
       in ${\rm H\beta}$ for optically thick nebulae.}
    \label{dqoEWhbmc}
\end{figure}

Figures~\ref{dqoOHmc} and \ref{dqoEWhbmc} represent
${\rm \Delta \log Q(H^0)}$ as a function of oxygen abundance
and ${\rm H\beta}$ equivalent width, respectively.
If the population synthesis process were exact, the recovered number
of ionizing photons would be equal to the true number of ionizing photons, and
in both cases ${\rm \Delta \log Q(H^0)}$ would be constant, equal to 0.
The only thing we measure here is the ability of the population synthesis
process to recover the number of ionizing photons. As can be seen in
Figs.~\ref{dqoOHmc} and \ref{dqoEWhbmc},
the dispersion obtained is equivalent to the dispersion observed in
Figs.~\ref{dqoOH} and \ref{dqoEWhb}; all the scatter in
${\rm \Delta \log Q(H^0)}$ can simply be
explained by the uncertainty introduced by the population synthesis process.
However, as can also be seen in Figs.~\ref{dqoOHmc}
and \ref{dqoEWhbmc}, the population synthesis process
does not introduce any systematical bias: most values of
${\rm \Delta \log Q(H^0)}$ lie around 0, and the distribution is fairly
symmetric. The systematical clustering of
the values of ${\rm \Delta \log Q(H^0)}$ around 0.8 cannot be explained
simply by an effect of the population synthesis process, and must be real.

\section{Summary}

For a sample of HII galaxies, we compared the number of predicted
ionizing photons from the population synthesis of their spectra
in the visible to the number of recombinations  counted in
the observed ${\rm H\beta}$.

Differences between these two numbers can arise (i) because the
evaluation of the number of ionizing photons is erroneous, either because
the theoretical ionizing spectra are wrong or because the
population synthesis algorithm is not able to evaluate the ionizing
spectrum correctly, or (ii) because a fraction of the ionizing photons escape
since the nebulae is optically thin.

We checked for systematic tendencies depending on metallicity (as measured
by the oxygen abundance from the gas) that would be indicative of
systematic deviations of the theoretical spectra in the EUV, as far as
the number of hydrogen ionizing photons is concerned.
We checked this with three different population synthesis libraries, and we
did not find any systematic effects for our sample.

Using Monte-Carlo simulations, we checked for the ability of our
population synthesis method to recover the
actual number of ionizing photons. Though most cases provided 
great scatter, we did not detect any systematic trend.

Finally, we examined the behaviour of the ratio of the number of predicted
ionizing photons to the number of recombinations as a function of
${\rm H\beta}$ equivalent width.
Though affected by the presence of older populations, the equivalent width
 in ${\rm H\beta}$ can be considered a gross indicator of
the nebular age, decreasing as the HII nebula becomes older.
We found a strong tendency for an increasing number of ionizing photons
escape the nebulae as the nebulae get older.

We interpret this as a consequence of the geometrical evolution of the
HII regions, which should  be optically thick to the
ionizing photons in some directions and optically thin in others.
We can understand it in terms
of a covering factor $f_c$ of the ionized gas that should decrease with time.
In other words, HII nebulae subfragment more and more
as they expand, always letting more ionizing photons escape.

The  absorption of ionizing photons by dust would provide an alternative
explanation, but we do not favour it, as it would require dust to absorb 
an always increasing fraction of  ionizing photons during the  evolution of   HII
regions. 

\begin{acknowledgements}
      We would like to thank FAPERJ, CNPq, and the Swiss National
      Foundation for financial support.
      F. C. wishes to thank the Astronomisches Institut der Universit\"at
      Basel for hospitality during a stay during which most of this paper
      was written.
      We wish to thank the anonymous referee for very helpful comments that
      helped to clarify and expand the discussion.    
\end{acknowledgements}

\appendix

\section{H$_{\beta}$ equivalent widths and population parameters of individual spectra}

\begin{table*}
   \begin{center}
      \caption{H$_{\beta}$ equivalent widths,  ${\rm \Delta \log Q(H^0)}$ parameters, mass ratios
                   (relative to the total stellar mass), and ages of the individual sub-populations
                     for each spectrum.}
      \label{EWparameters}
         \begin{tabular}{lrrrrrrr}
            \hline
     Spectra  & EW(H$_{\beta}$) & ${\rm \Delta \log Q(H^0)}$ & $M_{y}/M_{tot}$ & $Age_{y}$ & $M_{i}/M_{tot}$ & $Age_{i}$ & $M_{o}/M_{tot}$ \\
            \hline
  Cam08-28A(Cent)       & 117.4 & -2.688 &       0 &         - &  0.9091 & 2.026e+08 &  0.0909 \\
  Cam08-28A(E)          & 205.1 & -2.840 &       0 &         - &       1 & 2.026e+08 &       0 \\
  Cam0840+1044          & 248.5 & -0.997 &  0.0081 &  9.12e+06 &  0.2419 & 2.026e+08 &    0.75 \\
      CTS1006           & 351.2 & -0.988 &  0.0029 & 6.918e+06 &   0.088 & 2.026e+08 &  0.9091 \\
      CTS1008           & 646.5 &  0.410 &  0.0096 &     1e+06 &  0.0003 & 5.088e+08 &  0.9901 \\
      CTS1011           & 409.1 &  0.588 &  0.0099 & 1.995e+06 &       0 &         - &  0.9901 \\
      CTS1013           & 129.8 &  0.679 &  0.0029 &     1e+06 &  0.0293 & 5.088e+08 &  0.9677 \\
      CTS1016           & 126.9 &  0.952 &  0.0083 &     1e+06 &  0.0826 & 2.026e+08 &  0.9091 \\
      CTS1018           & 309.6 &  0.859 &  0.0161 &     1e+06 &  0.0161 & 5.088e+08 &  0.9677 \\
      CTS1019           &   423 & -0.221 &  0.0025 & 3.981e+06 &  0.0074 &     5e+07 &  0.9901 \\
      CTS1028           & 344.4 &  0.588 &  0.0227 &     1e+06 &  0.0682 &     5e+07 &  0.9091 \\
      CTS1033           & 259.4 &  0.409 &  0.0029 &     1e+06 &  0.0293 & 2.026e+08 &  0.9677 \\
      CTS1034           & 94.08 &  0.859 &  0.0029 & 1.995e+06 &  0.0293 & 1.015e+08 &  0.9677 \\
   DDO060(NW)           & 30.67 & -1.637 &       0 &         - &  0.0909 & 2.026e+08 &  0.9091 \\
   DDO060(SE)           & 38.55 & -1.751 &       0 &         - &  0.0909 & 2.026e+08 &  0.9091 \\
    DDO070(E)           & 225.3 &  0.796 &  0.0081 &     1e+06 &  0.0242 & 5.088e+08 &  0.9677 \\
    DDO070(W)           & 210.9 &  0.741 &  0.0081 &     1e+06 &  0.0242 & 2.026e+08 &  0.9677 \\
  DDO075(CentSW)        & 81.24 &  0.831 &  0.0161 & 5.012e+06 &  0.0161 & 1.995e+07 &  0.9677 \\
   DDO155(NE)           & 75.51 & -0.506 &  0.0029 & 6.918e+06 &   0.088 & 1.015e+08 &  0.9091 \\
   DDO155(SW)           & 102.8 &  1.037 &  0.0293 &     1e+06 &  0.8798 & 5.088e+08 &  0.0909 \\
  ESO289IG037           & 54.44 & -0.319 &  0.0029 & 7.943e+06 &  0.0293 & 1.015e+08 &  0.9677 \\
   ESO533G014           & 32.28 & -1.648 &       0 &         - &  0.0909 & 2.026e+08 &  0.9091 \\
    Fairall30           & 383.5 &  0.556 &   0.009 &     1e+06 &  0.0009 & 5.088e+08 &  0.9901 \\
       Haro24           &  74.1 & -2.069 &       0 &         - &    0.25 & 2.026e+08 &    0.75 \\
    IC5154(N)           & 40.17 & -0.162 &  0.0083 &  9.12e+06 &  0.0826 & 2.026e+08 &  0.9091 \\
    IC5154(S)           & 53.58 & -0.495 &   0.001 & 6.918e+06 &  0.0312 & 1.015e+08 &  0.9677 \\
  Marseille01           & 144.7 & -0.541 &  0.0081 &  9.12e+06 &  0.0242 &     5e+07 &  0.9677 \\
  Marseille68           & 114.3 &  0.893 &  0.0025 & 1.995e+06 &  0.0074 & 2.026e+08 &  0.9901 \\
  MBG02411-1457         & 13.98 & -1.261 &       0 &         - &  0.0909 & 5.088e+08 &  0.9091 \\
  MBG21567-1645         & 17.91 & -1.560 &       0 &         - &  0.0323 &     5e+07 &  0.9677 \\
  MBG22012-1550(E)      & 62.18 & -0.409 &  0.0081 &  9.12e+06 &  0.2419 & 2.026e+08 &    0.75 \\
        Mrk36           & 321.3 &  0.727 &  0.0227 &     1e+06 &  0.0682 &     5e+07 &  0.9091 \\
       Mrk710           & 170.9 &  0.937 &  0.0081 &     1e+06 &  0.0242 &     5e+07 &  0.9677 \\
  Mrk710(CentNE)        & 143.6 & -2.294 &       0 &         - &  0.0909 & 1.015e+08 &  0.9091 \\
   Mrk710(NE)           & 185.1 &  0.857 &  0.0083 &     1e+06 &  0.0826 & 2.026e+08 &  0.9091 \\
       Mrk711           & 135.9 & -0.065 &  0.0003 &     1e+06 &  0.0096 & 1.015e+08 &  0.9901 \\
      Mrk1318           & 323.9 &  0.253 &  0.0029 & 1.995e+06 &  0.0293 &     5e+07 &  0.9677 \\
      NGC7323           & 20.61 &  0.366 &       0 &         - &  0.0323 &     1e+07 &  0.9677 \\
   NGC7323(E)           & 7.079 &  0.337 &  0.0003 & 5.012e+06 &  0.0096 &     1e+07 &  0.9901 \\
   NGC7323(W)           & 13.28 &  0.356 &  0.0003 & 5.012e+06 &  0.0096 &     1e+07 &  0.9901 \\
  Tol0104-388(NW)       & 267.6 &  0.930 &  0.0161 & 1.995e+06 &  0.0161 & 5.088e+08 &  0.9677 \\
  Tol0104-388(SE)       & 14.71 &  0.218 &       0 &         - &       1 & 5.088e+08 &       0 \\
  Tol0117-414EW         &  66.1 & -0.402 &  0.0081 &  9.12e+06 &  0.2419 & 5.088e+08 &    0.75 \\
  Tol0117-414NS(CentN)  & 42.38 & -0.072 &  0.0029 & 6.918e+06 &   0.088 & 2.026e+08 &  0.9091 \\
  Tol0117-414NS(CentS)  &  77.4 &  0.907 &  0.0029 & 1.995e+06 &  0.0293 & 1.015e+08 &  0.9677 \\
  Tol0117-414NS(N)      & 158.1 & -0.308 &  0.0161 & 7.943e+06 &  0.0161 & 1.995e+07 &  0.9677 \\
  Tol0226-390           &   439 &  0.324 &  0.0081 &     1e+06 &  0.0242 & 1.995e+07 &  0.9677 \\
  Tol0306-405           & 228.3 &  0.749 &  0.0161 &     1e+06 &  0.0161 & 5.088e+08 &  0.9677 \\
  Tol0341-407(E)        & 167.3 &  0.774 &  0.0081 &     1e+06 &  0.0242 &     5e+07 &  0.9677 \\
  Tol0341-407(W)        & 217.2 & -1.759 &       0 &         - &  0.0323 & 1.995e+07 &  0.9677 \\
  Tol0440-381           & 143.8 &  1.194 &  0.0323 & 1.995e+06 &       0 &         - &  0.9677 \\
  Tol0528-383(E)        & 102.9 & -0.438 &  0.0081 & 7.943e+06 &  0.2419 & 5.088e+08 &    0.75 \\
  Tol0528-383(W)        & 160.4 & -0.679 &  0.0293 &  9.12e+06 &  0.8798 & 5.088e+08 &  0.0909 \\
            \hline
         \end{tabular}
   \end{center}
   \end{table*}
\addtocounter{table}{-1}
   \begin{table*}
   \begin{center}
      \caption{H$_{\beta}$ equivalent widths,  ${\rm \Delta \log Q(H^0)}$ parameters, mass ratios
                   (relative to the total stellar mass), and ages of the individual sub-populations
                     for each spectrum (continued).}
         \begin{tabular}{lrrrrrrr}
            \hline
     Spectra  & EW(H$_{\beta}$) & ${\rm \Delta \log Q(H^0)}$ & $M_{y}/M_{tot}$ & $Age_{y}$ & $M_{i}/M_{tot}$ & $Age_{i}
$ & $M_{o}/M_{tot}$ \\
            \hline
  Tol0610-387           & 29.41 & -2.003 &       0 &         - &    0.25 & 2.026e+08 &    0.75 \\
  Tol0645-376           &   122 &  0.696 &  0.0029 & 1.995e+06 &  0.0293 &     5e+07 &  0.9677 \\
  Tol0957-278(NW)       &   267 &  0.662 &  0.0081 & 1.995e+06 &  0.0242 &     5e+07 &  0.9677 \\
  Tol0957-278(SE)       & 186.5 &  1.016 &  0.0161 &     1e+06 &  0.0161 & 2.026e+08 &  0.9677 \\
  Tol1004-296(NW)       & 330.9 &  0.683 &  0.0074 &     1e+06 &  0.0025 & 2.026e+08 &  0.9901 \\
  Tol1004-296(SE)       & 306.3 &  0.669 &  0.0081 & 1.995e+06 &  0.0242 & 1.015e+08 &  0.9677 \\
  Tol1025-285           & 53.61 &  0.979 &  0.0029 & 1.995e+06 &   0.088 & 2.026e+08 &  0.9091 \\
  Tol1147-283           & 168.9 & -0.349 &  0.0025 & 6.026e+06 &  0.0074 & 1.995e+07 &  0.9901 \\
  Tol1345-420           & 226.8 &  0.460 &  0.0029 &     1e+06 &  0.0293 & 2.026e+08 &  0.9677 \\
  Tol1457-262E          & 413.2 &  0.637 &  0.0099 &     1e+06 &       0 &         - &  0.9901 \\
  Tol1457-262W(Cent)    & 146.3 &  0.645 &  0.0029 & 1.995e+06 &  0.0293 & 1.015e+08 &  0.9677 \\
  Tol1457-262W(CentE)   & 180.9 &  0.887 &  0.0083 &     1e+06 &  0.0826 & 5.088e+08 &  0.9091 \\
  Tol1457-262W(CentW)   & 236.7 &  0.717 &  0.0081 &     1e+06 &  0.0242 & 1.995e+07 &  0.9677 \\
  Tol1457-262W(E)       & 204.5 &  0.716 &  0.0081 &     1e+06 &  0.2419 & 5.088e+08 &    0.75 \\
  Tol1457-262W(W)       & 215.4 &  0.993 &  0.0161 &     1e+06 &  0.0161 & 5.088e+08 &  0.9677 \\
  Tol1924-416(E)        & 541.3 &  0.642 &  0.0682 & 1.995e+06 &  0.0227 &     1e+07 &  0.9091 \\
  Tol1924-416(W)        & 497.2 &  0.606 &  0.0161 &     1e+06 &  0.0161 &     5e+07 &  0.9677 \\
  Tol1937-423           & 31.84 & -0.375 &  0.0029 &  9.12e+06 &   0.088 & 2.026e+08 &  0.9091 \\
   UM69(Cent)           & 46.25 &  0.186 &  0.0227 &  9.12e+06 &  0.2273 & 5.088e+08 &    0.75 \\
      UM69(W)           & 91.47 & -1.153 &       0 &         - &     0.5 & 1.995e+07 &     0.5 \\
  UM137(CentE)          & 18.56 & -1.559 &       0 &         - &     0.5 & 5.088e+08 &     0.5 \\
  UM137(CentW)          & 24.19 & -1.793 &       0 &         - &     0.5 & 5.088e+08 &     0.5 \\
     UM137(W)           & 19.76 &  1.309 &  0.0029 &     1e+06 &   0.088 & 5.088e+08 &  0.9091 \\
  UM160(Cent)           &   137 &  0.409 &  0.0161 & 3.981e+06 &  0.4839 & 2.026e+08 &     0.5 \\
     UM160(E)           & 242.4 & -0.053 &  0.0081 & 5.012e+06 &  0.0242 &     1e+07 &  0.9677 \\
     UM160(W)           & 150.6 &  0.277 &  0.0083 & 5.012e+06 &  0.0826 & 5.088e+08 &  0.9091 \\
        UM191           & 37.62 &  0.056 &  0.0083 &  9.12e+06 &  0.0826 & 5.088e+08 &  0.9091 \\
        UM238           & 135.8 & -0.407 &  0.0025 & 6.918e+06 &  0.0074 &     1e+07 &  0.9901 \\
        UM307           & 118.1 &  0.011 &  0.0081 & 6.026e+06 &  0.2419 & 5.088e+08 &    0.75 \\
        UM323           & 113.6 &  0.659 &       1 & 5.012e+06 &       0 &         - &       0 \\
        UM395           & 35.77 &  0.154 &  0.0081 & 7.943e+06 &  0.2419 & 5.088e+08 &    0.75 \\
        UM396           & 501.3 &  0.484 &  0.0161 &     1e+06 &  0.0161 & 1.015e+08 &  0.9677 \\
        UM408           & 156.1 &  0.942 &  0.0081 & 1.995e+06 &  0.0242 & 1.015e+08 &  0.9677 \\
  UM439(Cent)           & 110.2 &  0.979 &  0.0083 & 1.995e+06 &  0.0826 & 2.026e+08 &  0.9091 \\
    UM439(NW)           & 223.3 &  0.759 &  0.0081 & 1.995e+06 &  0.0242 &     5e+07 &  0.9677 \\
  UM448(extensionNE-SW) & 242.6 &  0.400 &  0.0029 & 1.995e+06 &  0.0293 &     5e+07 &  0.9677 \\
    UM455(NW)           & 190.2 &  0.554 &  0.0029 & 1.995e+06 &  0.0293 & 2.026e+08 &  0.9677 \\
  UM456(Cent)           & 239.9 &  0.721 &  0.0227 &     1e+06 &  0.2273 & 1.015e+08 &    0.75 \\
    UM456(NE)           & 403.2 &  0.720 &  0.0682 &     1e+06 &  0.6818 & 5.088e+08 &    0.25 \\
    UM456(SW)           & 215.5 &  0.840 &  0.0227 &     1e+06 &  0.0682 & 1.995e+07 &  0.9091 \\
     UM461(E)           & 854.2 &  0.448 &  0.0323 &     1e+06 &       0 &         - &  0.9677 \\
     UM461(W)           & 638.4 &  0.598 &  0.0682 &     1e+06 &  0.0227 & 5.088e+08 &  0.9091 \\
    UM462(NE)           & 301.5 &  0.785 &  0.0455 &     1e+06 &  0.0455 & 1.995e+07 &  0.9091 \\
    UM462(SW)           & 382.4 & -0.158 &   0.125 & 5.012e+06 &   0.375 &     1e+07 &     0.5 \\
     UM499(E)           & 145.4 & -0.821 &  0.0029 &  9.12e+06 &  0.0293 & 2.026e+08 &  0.9677 \\
     UM499(W)           & 58.06 & -0.529 &  0.0029 &  9.12e+06 &   0.088 & 2.026e+08 &  0.9091 \\
     UM533(E)           & 138.4 & -0.807 &  0.0161 & 7.943e+06 &  0.4839 & 5.088e+08 &     0.5 \\
  UM559(Cent)           & 33.93 &  0.236 &  0.0081 & 7.943e+06 &  0.2419 & 2.026e+08 &    0.75 \\
  UM598W(CentNE)        &  20.3 & -1.289 &       0 &         - &  0.0323 & 2.026e+08 &  0.9677 \\
  UM598W(CentSW)        & 46.34 &  1.142 &  0.0025 & 1.995e+06 &  0.0074 & 2.026e+08 &  0.9901 \\
   UM598W(NE)           & 58.18 &  0.595 &   0.001 &     1e+06 &  0.0312 &     5e+07 &  0.9677 \\
   UM598W(SW)           & 112.8 &  1.029 &  0.0081 &     1e+06 &  0.2419 & 5.088e+08 &    0.75 \\
            \hline
         \end{tabular}
   \end{center}
   \end{table*}
%
%

\end{document}